\documentstyle[12pt,epsfig]{article}

\newcommand{\be}{\begin{equation}}
\newcommand{\ee}{\end{equation}}
\begin{document}
\topmargin 0pt
\oddsidemargin=-0.4truecm
\evensidemargin=-0.4truecm
\renewcommand{\thefootnote}{\fnsymbol{footnote}}
\newpage
\setcounter{page}{1}
\begin{titlepage}
\vspace*{-2.0cm}   
\begin{flushright}
TUM-HEP-423/01 \\
hep-th/0107223
\end{flushright}
\vspace*{0.5cm}
\begin{center}
\vspace*{0.2cm}  
{\Large \bf Dynamical localization of gauge fields on a brane}
\\
\vspace{1.0cm} 
{\large E. Kh. Akhmedov
\footnote{On leave from National Research Centre Kurchatov Institute,
Moscow 123182, Russia. E-mail address: akhmedov@physik.tu-muenchen.de}}\\
\vspace{0.2cm}
{\em Institut f{\"u}r Theoretische Physik, Physik Department,
Technische Universit{\"a}t M{\"u}nchen James-Franck-Stra{\ss}e,
D-85748 Garching b. M{\"u}nchen, Germany} 

\end{center}

\vglue 1.5truecm

\begin{abstract}
We propose a dynamical mechanism of localization of gauge fields on a brane 
in which gauge bosons are excitations of the brane itself or composites made 
out of matter fields localized on the brane. The mechanism is operative for 
both Abelian and non-Abelian gauge fields. Several scalar and scalar-fermion 
composite models of gauge fields are considered. The models exhibit exact 
gauge invariance and therefore charge universality of gauge interactions is 
automatically preserved. The mechanism is shown to be equivalent to a 
modification of the Dvali, Gabadadze and Shifman scenario in which gauge 
bosons have no bulk kinetic terms and only possess induced kinetic terms on 
the brane.  
\end{abstract}

\vspace{1.cm}
\vspace{.3cm}
\end{titlepage}
\renewcommand{\thefootnote}{\arabic{footnote}}
\setcounter{footnote}{0}
\newpage
\section{Introduction}

In the brane world scenarios with large or infinite extra dimensions 
[1 -- 8] one has to explain why we live on the
brane and do not escape into extra dimensions. In other words, one needs a 
mechanism by which the ordinary (standard-model) matter is trapped on the 
brane while only gravity and possibly some other particles which are the
singlets of the standard model can propagate in the bulk.  

While simple mechanisms of trapping scalars and fermions have been 
constructed \cite{RS,A,V}, localizing gauge fields on the brane is 
notoriously difficult. The main problem turned out to be preserving 
charge universality of gauge interactions. The interactions of the localized 
charged particles with gauge fields depend in general not only on their
charges but also on their wave functions in the directions transverse to the 
brane, thus violating charge universality \cite{R,DR}. Several mechanisms 
of localization of gauge fields on a brane which evade this difficulty 
have been suggested so far, both in flat \cite{DS} and warped \cite{O,DRT}
space-times. Nongravitational mechanisms are of particular interest as in
some popular brane world scenarios extra dimensions are flat 
\cite{RS,ADD1,ADD2}. 

In the present Letter we propose a simple mechanism
of localization of gauge fields on a brane which does not rely on gravity 
and so can work in both flat and warped space-times. It is operative for 
Abelian as well as non-Abelian gauge fields. In our mechanism 
gauge fields are composites made out of localized scalar or fermion fields. 
The localized matter fields can be either fluctuations of the brane itself, 
in which case the composite gauge fields are the massless vector excitations 
of the brane, or they can be other zero-mode scalar or fermion fields 
localized on the brane. We show that in pure fermionic composite models of
gauge fields gauge invariance cannot be naturally implemented, while
scalar and scalar-fermion models can be made gauge invariant, thus 
preserving charge universality automatically. We demonstrate how the
higher-dimensional gauge invariance translates into the exact gauge
invariance of the effective four-dimensional theory irrespective of the 
shapes of the localization wave functions of the matter fields. 
 
We also show that our mechanism is formally equivalent to a modification
of the Dvali, Gabadadze and Shifman (DGS) scenario \cite{DGS}. In \cite{DGS} 
a mechanism of quasi-localization of gauge fields on a brane was proposed, 
in which gauge fields, in addition to bulk kinetic terms, have induced 
kinetic terms on the brane. The gauge fields are localized and their 
interactions are essentially four-dimensional at distances small compared
to a crossover scale $r_c$, while at distances larger than $r_c$ gauge 
interactions are higher-dimensional and gauge fields can escape to the
bulk. It has been argued in \cite{DR} that, while this scenario is viable 
in a five-dimensional space-time, it may have problems when the number of 
extra dimensions $d\ge 2$: 
in the case of the $\delta$-function type brane the gauge boson propagator 
does not exist, while for finite-thickness branes charge universality 
cannot be preserved. Our mechanism is essentially equivalent to a 
modification of the DGS scenario in which gauge fields have only induced 
kinetic terms on the brane and no bulk kinetic terms. The mechanism thus 
leads to the exact localization of gauge fields on a brane rather than 
to quasi-localization. In addition, because of the absence of the gauge 
boson kinetic terms in the bulk, the propagators of the gauge bosons exist
and charge universality is preserved in space-times with an arbitrary  
number of extra dimensions $d$ and for both $\delta$-function type and 
finite-thickness branes.   

\section{Fermionic models}
Fermionic composite models of gauge fields have been widely discussed in
the literature. Most of them are based on the Bjorken model \cite{Bjor}
with the nonlinear Lagrangian 
\be
{\cal L}(\psi,\bar{\psi}) =
\bar{\psi}(i\partial\!\!\!/ - M)\psi
-G(\bar{\psi}\gamma^\mu\psi)(\bar{\psi}\gamma_\mu\psi)\,.
\label{L1f}
\ee
The standard technique of dealing with such a nonlinear model is to
linearize it by introducing an auxiliary vector field $A_\mu$. Indeed, 
the generating functional in the model 
\be
Z_1=\int {\cal D}\psi{\cal D}\bar{\psi}\,e^{i\!\int {\cal
L}(\psi,\bar{\psi})
d^4 x} 
\label{Z1}
\ee
can be rewritten as 
\be
Z_2=\int {\cal D}\psi{\cal D}\bar{\psi}{\cal D}A \,e^{i\!\int {\cal
L}(\psi,\bar{\psi},A)
d^4 x} 
\label{Z2}
\ee
where 
\be
{\cal L}(\psi,\bar{\psi},A) = \bar{\psi}(i\partial\!\!\!/ - M)\psi
-e_0 \bar{\psi}\gamma_\mu\psi A^\mu+\frac{m_0^2}{2} 
A_\mu A^\mu\,.
\label{L2f}
\ee
The path integral over $A_\mu$ in (\ref{Z2}) is Gaussian, and by performing 
it one recovers the generating functional $Z_1$ of eq. (\ref{Z1}) with the 
identification 
\be
G=\frac{e_0^2}{2m_0^2}\,.
\label{G}
\ee
The Lagrangian in eq. (\ref{L2f}) describes a spin-1/2 field interacting
with the vector field $A_\mu$. The theory is reminiscent of the spinor QED
except that the field $A_\mu$ has a mass term which breaks gauge invariance,  
and does not have a kinetic term. The non-propagating classical auxiliary
field $A_\mu$ acquires the kinetic term through quantum fluctuations 
of the fermion field and so becomes a physical propagating field 
\cite{Sakh,Zel}; at one fermion loop level one finds 
\be
{\cal L}_{kin}\simeq -\frac{e_0^2}{12\pi^2}\ln(\Lambda^2/M^2)\frac{1}{4}
F_{\mu\nu}F^{\mu\nu} \equiv -Z_3 \frac{1}{4}F_{\mu\nu}F^{\mu\nu} 
\label{Lkin1}
\ee
where $\Lambda$ is the ultraviolet cutoff, and after the renormalization 
$A_\mu\to\sqrt{Z_3}A_\mu$ one gets the standard spinor QED with a massive
photon field. 

The nonvanishing photon mass $m_0\ne 0$ in eq. (\ref{L2f}) is clearly related 
to the fact that the original Lagrangian (\ref{L1f}) is not in general gauge 
invariant. There have been several suggestions of how to deal with this 
problem. One possibility \cite{inf} is to consider the limit $m_0\to 0$
which through eq. (\ref{G}) is equivalent to $G\to\infty$. This, however, 
does not appear to be a satisfactory solution. It is easy to see that the 
Lagrangian (\ref{L1f}) indeed becomes gauge invariant in this limit, but 
at the expense of neglecting the gauge-noninvariant kinetic term compared
to the current-current term which has local $U(1)$ symmetry. This means 
that the fermionic field $\psi$ becomes an unphysical non-propagating field 
in this limit; in particular, the gauge boson kinetic term can no longer 
be generated through the fermion loops. An alternative suggestion \cite{zero} 
was to require that the current $j_\mu(x)=\bar{\psi}(x)\gamma_\mu\psi(x)$ 
vanish identically, which makes the kinetic term in the fermionic 
Lagrangian gauge invariant. However, in this case one obtains a 
non-interacting gauge boson field, which is not of much interest. 
Yet another possibility \cite{Eguchi} is to cancel the photon mass term
in (\ref{L2f}) against the gauge-noninvariant ${\cal O}(\Lambda^2)$
contribution coming from the one-fermion-loop self energy of photon 
calculated with the Euclidean
momentum cutoff. In this approach one considers the photon mass term as 
a counter term introduced to compensate for the use of a gauge-noninvariant 
regularization. This, however, seems to be rather artificial as gauge 
invariance does not follow from the form of the Lagrangian but is rather 
imposed on the theory ``by hand''.  In addition, the argument does not
apply if one employs a gauge-invariant regularization. 

To summarize, the fermionic models are not quite satisfactory as they have 
difficulties ensuring gauge invariance of the induced gauge boson 
theory. They may, however, be useful if one considers gauge invariance as 
an approximate symmetry valid only at distances small compared to the  
scale $R\sim m_0^{-1}$. It is not difficult to construct a higher-dimensional 
generalization of the Lagrangian (\ref{L1f}) with the fermionic chiral
zero mode $\Psi$ localized on a 3-dimensional brane. 
For example, in a five-dimensional space-time one can write 
\be
{\cal L}(\Psi,\bar{\Psi}) =
\bar{\Psi}i\Gamma^B\partial_B \Psi + \Delta{\cal L}
-G_{(5)}(\bar{\Psi}\Gamma^B\Psi)(\bar{\Psi}\Gamma_B\Psi)\,. 
\label{L3f}
\ee
Here $\Psi(x,z)=u(z)\psi(x)$, $x^\mu$ ($\mu=0,1,2,3$) and $z$ 
are the coordinates along the brane and in the transverse (fifth) direction  
respectively, $\Gamma_B$ ($B=0,1,2,3,5$) are the five-dimensional gamma 
matrices: $\Gamma_\mu=\gamma_\mu$, $\Gamma_5=-i\gamma_5$, and $\Delta{\cal L}$ 
describes the fermion-brane interaction. The localization 
wave function $u(z)$ falls off at the distances $|z|\sim m^{-1}$ where 
$m^{-1}$ is the brane thickness. The model can be linearized by introducing 
an auxiliary 5-vector field ${\cal A}_B=({\cal A}_\mu, {\cal A}_5)$. 
At the one fermion loop level the field ${\cal A}_B$ acquires a
gauge-invariant kinetic term which is localized on the brane because 
the fermions are trapped there. 

The model sketched above is not in general gauge invariant and therefore
may have problems ensuring charge universality of gauge interactions. 
We therefore will concentrate on scalar and scalar-fermionic models in 
which exact gauge invariance can be naturally implemented. 

\section{Scalar and scalar-fermion models in four dimensions}
The origin of gauge-noninvariance of the pure fermionic models discussed
above can be traced back to the quadratic in $A_\mu$ terms in the auxiliary 
Lagrangians. Such quadratic terms are necessary for the path integrals
over $A_\mu$ to be Gaussian, and in fermionic theories they are nothing
but the mass terms of the auxiliary vector fields which break gauge 
invariance. In contrast to this, in scalar theories 
$A_\mu^2$ terms do not in general break gauge invariance; moreover, such 
terms are actually necessary to ensure gauge invariance.

We shall consider the nonlinear scalar model with the Lagrangian 
\be
{\cal L}(\phi,\phi^\dag) = 
\partial_\mu \phi^\dag \partial^\mu \phi -V(\phi^\dag\phi)-\frac{
(i\phi^\dag\!\buildrel\leftrightarrow\over {\partial^\mu}\!\phi)
(i\phi^\dag\!\buildrel\leftrightarrow\over {\partial_\mu}\!\phi)}
{4\phi^\dag\phi}\,.
\label{L1s}
\ee
This Lagrangian is invariant with respect to the local $U(1)$ transformation 
$\phi\to e^{i\alpha(x)}\phi$ despite the absence of the gauge fields 
\footnote{Scalar theories possessing local gauge invariance and
generating gauge fields dynamically have been previously discussed in the 
framework of the non-linear sigma model (see, e.g., \cite{nonlin}). In
contrast to these models, we do not impose any constraints on the scalar 
field $\phi$ in (\ref{L1s}).}. The model can be linearized with the help
of the auxiliary vector field $A_\mu$, the Lagrangian of the model being 
\be
{\cal L}(\phi,\phi^\dag, A) = 
\partial_\mu \phi^\dag \partial^\mu \phi -V(\phi^\dag\phi)-
e_0 (i\phi^\dag\!\buildrel\leftrightarrow\over {\partial_\mu}\!\phi) A^\mu 
+e_0^2\,\phi^\dag \phi\,A_\mu A^\mu\,.
\label{L2s}
\ee
The last (quadratic in $A_\mu$) term is not an $A_\mu$ mass term but
rather is the $\phi\phi A A$ coupling which is required by gauge 
invariance. Integrating over $A_\mu$ in the path integral one arrives at
the generating functional of the model (\ref{L1s}). 

The Lagrangian (\ref{L2s}) describes scalar QED without the kinetic term of 
the photon field. At the classical level the equation of motion of $A_\mu$
expresses it in terms of the scalar field: 
\be
A_\mu=\frac{1}{2e_0}\frac{i\phi^\dag\!\buildrel\leftrightarrow\over
{\partial_\mu}\!\phi}{\phi^\dag\phi}\,.
\label{A1}
\ee
The field (\ref{A1}) has the correct transformation properties under the 
$U(1)$ gauge transformation, $A_\mu\to A_\mu-(1/e_0)\,\partial_\mu\alpha(x)$   
(notice that this is not so in the fermionic case). 
Quantum fluctuations of the scalar field induce the usual gauge-invariant
kinetic term for $A_\mu$. At one loop level two diagrams contribute,
yielding 
\be
{\cal L}_{kin}\simeq -\frac{e_0^2}{48\pi^2}\ln(\Lambda^2/\mu^2)\frac{1}{4}
F_{\mu\nu}F^{\mu\nu} \equiv -Z_3 \frac{1}{4}F_{\mu\nu}F^{\mu\nu} 
\label{Lkin2}
\ee
where $\Lambda$ and $\mu$ are the ultraviolet and infrared cutoffs,
respectively. After the renormalization $A_\mu\to\sqrt{Z_3}A_\mu$ one
obtains the standard Lagrangian of scalar QED. Notice that the 
renormalized parameters do not depend on the redundant parameter $e_0$, 
the renormalized charge being 
\be
e^2(\mu)=\frac{48\pi^2}{\ln(\Lambda^2/\mu^2)}\,.
\label{e2}
\ee
The fact that there is no charge parameter in the original Lagrangian 
(\ref{L1s}) and the physical charge is generated dynamically is related to
the circumstance that the kinetic term of the gauge field is generated
dynamically. 

A comment is in order at this point. In eqs. (\ref{Lkin1}) and
(\ref{Lkin2})
and in similar formulas below we neglect the terms of order unity as well
as terms containing positive powers of $p^2/\Lambda^2$ (where $p$ is an 
external momentum which we assume to be small compared to $\Lambda$) and
only 
retain $\log\Lambda$ terms. While the logarithmic terms are universal, 
${\cal O}(1)$ and smaller terms depend on the details of the regularization 
scheme used and, with the ultraviolet cutoff $\Lambda$ in place, even on the 
momentum routing along the loops. These model-dependent terms can be
neglected if $\log\Lambda$ terms are large, which we assume.  

We have demonstrated that physical gauge bosons can be generated dynamically 
in nonlinear scalar models with Lagrangians of the type
(\ref{L1s}). Several questions then naturally arise:

$\bullet$ Can the model be generalized to the case of several scalar
fields with different charges?

$\bullet$ Can charged fermions be incorporated in this scenario?

$\bullet$ Can non-Abelian gauge fields be generated in a similar way?

We shall now answer these questions in turn. 

Assume that we have $n$ scalar fields with the charges $e_i$ assembled
into a vector $\phi=(\phi_1,...,\phi_n)$. The $U(1)$ gauge transformation 
for $\phi$ is $\phi\to e^{iq\alpha(x)}\phi$, where $q$ is the matrix  
of the charges. It is then easy to see that the Lagrangian 
\be
{\cal L}(\phi,\phi^\dag) = 
\partial_\mu \phi^\dag \partial^\mu \phi -V(\phi^\dag\phi)-\frac{
(i\phi^\dag q\!\buildrel\leftrightarrow\over{\partial^\mu}\!\phi)
(i\phi^\dag q\!\buildrel\leftrightarrow\over{\partial_\mu}\!\phi)}
{4\,\phi^\dag q^2\phi}
\label{L3s}
\ee
has the local $U(1)$ symmetry. The application of the auxiliary field 
formalism is then straightforward; the model is equivalent to the 
usual QED of $n$ charged scalar fields. 

Once the model contains scalars so that the $A_\mu^2$ terms in the
auxiliary Lagrangians are gauge invariant, one can easily incorporate 
fermions. For example, in the case of one scalar and one spinor field 
the nonlinear Lagrangian of the model is 
\be
{\cal L}(\phi,\phi^\dag,\psi,\bar{\psi}) = 
\partial_\mu \phi^\dag \partial^\mu \phi -V(\phi^\dag\phi)
+\bar{\psi}(i\partial\!\!\!/ - M)\psi
-\frac{(i\phi^\dag \!\buildrel\leftrightarrow\over{\partial_\mu}\!\phi 
+\bar{\psi}\gamma_\mu\psi)^2}{4\phi^\dag \phi}
\label{L1sf}
\ee
One can readily make sure that it is gauge invariant. The auxiliary vector
field is introduced in the usual way. At the classical level, its equation 
of motion expresses it through the scalar and spinor fields: 
\be
A_\mu=\frac{1}{2e_0}\frac{(i\phi^\dag\!\buildrel\leftrightarrow\over
{\partial_\mu}\!\phi+\bar{\psi}\gamma_\mu \psi)}{\phi^\dag\phi}\,.
\label{A2}
\ee
This field has the correct transformation properties under the 
$U(1)$ gauge transformation. 
The field $A_\mu$ becomes a physical propagating photon field after its 
kinetic term is induced by scalar and fermion loops. The resulting theory 
is the QED with scalar and spinor fields. It is easy to generalize the above 
model to the case of an arbitrary number of scalar and fermion fields with in 
general different electric charges. 

The mechanism under discussion can be used to generate non-Abelian
composite gauge fields as well. Consider the $SU(2)$ case as an 
example. Let the scalar field $\phi$ be in the fundamental representation;
then the Lagrangian 
\be
{\cal L}(\phi,\phi^\dag) = 
\partial_\mu \phi^\dag \partial^\mu \phi -V(\phi^\dag\phi)-\frac{
(i\phi^\dag\mbox{\boldmath$\tau$}\!\buildrel\leftrightarrow\over{\partial^\mu}
\!\phi)(i\phi^\dag \mbox{\boldmath$\tau$}\!\buildrel\leftrightarrow
\over{\partial_\mu}\!\phi)}{4\phi^\dag\phi}
\label{L4s}
\ee
possesses the local $SU(2)$ symmetry. It can be linearized with the auxiliary 
vector field $A_\mu^i$ in the adjoint representation. Since vector-scalar 
interactions are gauge invariant, the full gauge invariant kinetic term for 
$A_\mu^i$ is induced through one-scalar-loop diagrams; triple and quartic 
couplings come from the three-point and four-point $A_\mu^i$ functions, 
respectively. These functions have the same logarithmic renormalization 
factor $Z=Z_3$. One also obtains higher-dimension terms through one-loop 
diagrams with more than four external $A_\mu^i$ legs; however, it is easy to 
see that all these terms are either of order unity or contain positive powers 
of $p^2/\Lambda^2$ and so we neglect them.

\section{Higher-dimensional models}

We shall now consider higher-dimensional composite gauge boson models and 
discuss the localization of the gauge fields on three-dimensional branes. 
As our mechanism does not rely on gravity, we consider flat space-times. 
For definiteness, we consider models in $4+1$ dimensions; the generalization 
to the case of $d>1$ extra dimensions is straightforward.

We start with the case of a single scalar zero mode $\Phi$ localized on a
brane. The field $\Phi$ can be either a small fluctuation of the brane 
itself \cite{RS}, or an independent localized scalar field. The Lagrangian
of the model is 
\be
{\cal L}_{(5)}(\Phi,\Phi^\dag) =
\partial^B \Phi^\dag \partial_B \Phi +\Delta{\cal L}-\frac{
(i\Phi^\dag\!\buildrel\leftrightarrow\over {\partial^B}\!\Phi)
(i\Phi^\dag\!\buildrel\leftrightarrow\over {\partial_B}\!\Phi)}
{4\Phi^\dag\Phi}\,.
\label{L51}
\ee
Here $\Phi(x,z)=\varphi(z)\phi(x)$. The (real) localization wave function
$\varphi(z)$ falls off at the distances $|z|\sim m^{-1}$ from the brane, 
$m^{-1}$ being the brane thickness. It is normalized by the condition
\be
\int_{-\infty}^{\infty} dz\,\varphi^2(z)=1\,.
\label{norm}
\ee
The term $\Delta{\cal L}$ describes the interaction of the zero mode 
$\Phi$ with the brane; it cancels the term $\sim m^2 \varphi^2(z)\phi^\dag
\phi$ coming from the derivative over $z$ in the kinetic term: 
\footnote{For example, if the brane is described by a kink $\Phi_0(z)=
(m/\sqrt{\lambda})\tanh(m z/\sqrt{2})$ \cite{RS}, the linearized
equation of motion for small fluctuations $\Phi(x,z)$ of the brane,  
$[\partial^B \partial_B-m^2+3\lambda\Phi_0(z)^2]\Phi=0$, has a 
localized zero-mode solution $\Phi(x,z)=\varphi(z)\phi(x)$ with the 
normalized localization wave function $\varphi(z)=(3m)^{1/2}/2^{5/4}
\cosh^{-2}(m z/\sqrt{2})$. From the equation of motion for $\Phi(x,z)$ 
one reconstructs $\Delta{\cal L}=(-m^2+3\lambda\Phi_0(z)^2)\Phi^\dag
\Phi$, which leads to (\ref{compens}). Eq. (\ref{compens}) is, however,  
quite general and does not depend on the explicit form of the brane; 
it just reflects the fact that the localized field is a zero mode.}
\be
\partial^B \Phi^\dag \partial_B \Phi +\Delta{\cal L}=
\varphi^2(z)\partial^\mu\phi(x)^\dag\partial_\mu\phi(x)\,. 
\label{compens}
\ee
Since the localization wave function $\varphi(z)$ is real, one has 
\be
i\Phi^\dag\!\buildrel\leftrightarrow\over {\partial_B}\!\Phi=
\varphi^2(z)\,i\phi(x)^\dag\!\buildrel\leftrightarrow\over
{\partial_\mu}\!\phi(x)\,\delta_{B\mu}\,.
\label{current}
\ee
Putting eqs. (\ref{L51}), (\ref{compens}) and (\ref{current}) together 
we arrive at
\be
{\cal L}_{(5)} =\varphi^2(z)\left\{ 
\partial^\mu \phi^\dag \partial_\mu \phi -\frac{
(i\phi^\dag\!\buildrel\leftrightarrow\over {\partial^\mu}\!\phi)
(i\phi^\dag\!\buildrel\leftrightarrow\over {\partial_\mu}\!\phi)}
{4\phi^\dag\phi}\right\}\,.
\label{L52}
\ee
The effective four dimensional Lagrangian 
\be
{\cal L}_{(4)}=\int_{-\infty}^{\infty} \!dz\,{\cal L}_{(5)}
\label{4dim}
\ee
then coincides with the Lagrangian (\ref{L1s}) with $V(\phi^\dag \phi)=0$. 
One can now apply the auxiliary field formalism as discussed in detail in 
sec. 3 and show that the massless gauge boson field is produced, whose 
kinetic term is generated by scalar loops. Alternatively, one could apply 
the auxiliary field formalism already in the five-dimensional theory. 
The classical five-dimensional auxiliary field 
\be
{\cal A}_B=\frac{1}{2e_0}\frac{i\Phi^\dag\!\buildrel\leftrightarrow
\over{\partial_B}\!\Phi}{\Phi^\dag\Phi}=
\frac{1}{2e_0} \frac{i\phi^\dag\!\buildrel\leftrightarrow\over  
{\partial_\mu}\!\phi}{\phi^\dag\phi}\,\delta_{B\mu} = A_\mu 
\,\delta_{B\mu}
\label{A3}
\ee
does not depend on the transverse coordinate $z$ and so is not localized 
on the brane 
\footnote{This, in particular, means that the field (\ref{A3}) is not 
normalizable. This point, however, should be of no concern as the field
(\ref{A3}) is a non-propagating auxiliary field which simply gives an
alternative description of the scalar self-coupling. The integration of
the Lagrangian ${\cal L}_{(5)}$ of eq. (\ref{L52}) over $z$ does not lead 
to any divergence.}. 
At the same time, its loop-induced kinetic term is localized:
\be
{\cal L}_{kin(5)}=-\varphi^2(z)\, Z_3 \frac{1}{4}F_{\mu\nu}F^{\mu\nu}\,.
\label{Lkin3}
\ee
This means that the gauge boson field can only propagate on the brane.
Thus, our model has a gauge boson field which lives in the bulk but has
only an induced kinetic term on the brane. As was pointed out before, such 
a model is equivalent to a modification of the DGS scenario \cite{DGS}. 
The four-dimensional theory obtained after the integration over the fifth 
coordinate is identical to the theory resulting from the application of 
the auxiliary field formalism directly in the four-dimensional space-time.

As can be seen from eq. (\ref{L52}), in the case of one scalar field 
the $z$-dependence of ${\cal L}_{(5)}$ factorizes out. 
This, however, is not so if there are more than one scalar and/or
fermion fields with different localization wave functions. This raises a
question of how gauge invariance is preserved in the effective 
four-dimensional theory. Indeed, for gauge invariance to hold, the 
coefficients of different terms in ${\cal L}_{(4)}$ must have certain fixed 
relative values, while with arbitrary localization wave functions one can 
expect that upon the integration of ${\cal L}_{(5)}$ over $z$ these 
coefficients will take arbitrary values. We shall now show that in fact
this is not the case and demonstrate how the gauge invariance is actually
preserved in the four-dimensional theory. 

Consider first an example of one scalar and one spinor field with the 
localization wave functions $\varphi(z)$ and $u(z)$ respectively: $\Phi(x,z)=
\varphi(z)\phi(x)$, $\Psi(x,z)=u(z)\psi(x)$. We assume $\varphi(z)$ and $u(z)$ 
to be normalized according to (\ref{norm}). The Lagrangian of the model is 
\be
{\cal L}_{(5)}=\partial^B\Phi^\dag \partial_B\Phi+\bar{\Psi}
i\Gamma^B\partial_B\Psi+\Delta{\cal L}
-\frac{(i\Phi^\dag \!\buildrel\leftrightarrow\over{\partial_B}\!\Phi 
+\bar{\Psi}\Gamma_B\Psi)^2}{4\Phi^\dag \Phi}\,.
\label{L53} 
\ee
Here $\Delta{\cal L}$ is chosen in such a way that 
$$\partial^B\Phi^\dag\partial_B\Phi+\bar{\Psi}i\Gamma^B\partial_B\Psi+
\Delta{\cal L}=\varphi^2(z)\,\partial^\mu\phi^\dag\partial_\mu\phi+u^2(z)\,
\bar{\psi}i\gamma^\mu\partial_\mu\psi\,.$$
Thus one can rewrite eq. (\ref{L53}) as  
\be
{\cal L}_{(5)}=\varphi^2(z)\,\partial^\mu\phi^\dag\partial_\mu\phi+
u^2(z)\,\bar{\psi}i\gamma^\mu\partial_\mu\psi-\frac{[\varphi^2(z)\,i\phi^\dag
\!\buildrel\leftrightarrow\over{\partial_\mu}\!\phi 
+u^2(z)\,\bar{\psi}\gamma_\mu\psi]^2}{4\varphi^2(z)\phi^\dag \phi}\,,
\label{L54}
\ee
where we have used the fact that the localized fermionic zero modes are
chiral, so that $\bar{\Psi}\Gamma_5\Psi=\pm\bar{\Psi}\Psi=0$.
The last term in this expression is 
\be
-\frac{1}{4}\left[\varphi^2(z)\,\frac{(i\phi^\dag\!\buildrel\leftrightarrow
\over{\partial_\mu}\!\phi)^2}{\phi^\dag \phi}+
2u^2(z)\,\frac{(i\phi^\dag\!\buildrel\leftrightarrow\over{\partial_\mu}\!\phi)
\,\bar{\psi}\gamma^\mu\psi}{\phi^\dag \phi}+
\frac{u^4(z)}{\varphi^2(z)}\frac{(\bar{\psi}\gamma_\mu\psi)^2}
{\phi^\dag \phi}\right]\,.
\label{L55}
\ee
The integration of the first two terms in (\ref{L55}) over $z$ yields the
correct coefficients for these terms to produce, together with the 
(integrated) kinetic terms, a gauge invariant expression; the integral of
the last term is gauge invariant by itself. Thus we obtain 
\be
{\cal L}_{(4)} = 
\partial^\mu \phi^\dag \partial_\mu \phi +\bar{\psi}i\partial\!\!\!/\psi
-\frac{(i\phi^\dag \!\buildrel\leftrightarrow\over{\partial_\mu}\!\phi 
+\bar{\psi}\gamma_\mu\psi)^2}{4\phi^\dag\phi}+
C\,\frac{(\bar{\psi}\gamma_\mu\psi)^2}{\phi^\dag\phi}
\label{L56}
\ee
where
\be
C=-\frac{1}{4}\int_{-\infty}^{\infty}dz\frac{u^4(z)-\varphi^4(z)}
{\varphi^2(z)}\,.
\label{C}
\ee
Except for the last term, the Lagrangian in eq. (\ref{L56}) coincides with 
that in eq. (\ref{L1sf}) with $V(\phi^\dag\phi)=M=0$. The last term 
in (\ref{L56}) is a nonlinear gauge-invariant expression. Note that 
for $\varphi(z)=u(z)$ the constant $C$ vanishes; therefore when 
the localization wave functions of the spinor and scalar fields coincide, 
the theory is fully linearized by the dynamical generation of the gauge
boson. Otherwise the four-dimensional theory has a residual nonlinear 
coupling, even though the five-dimensional theory is fully linearized. 

Consider now a slightly more complicated case of two localized scalar fields 
with different localization wave functions, $\varphi_1(z)$ and $\varphi_2(z)$, 
both normalized according to (\ref{norm}). The five-dimensional Lagrangian
of the model is 
\be
{\cal L}_{(5)}=\sum_{i=1,2}\partial^B\Phi_i^\dag\partial_B\Phi_i+
\Delta{\cal L}-\frac{\left(i\sum\limits_{i=1,2}\Phi_i^\dag\!\buildrel
\leftrightarrow\over{\partial_B}\!\Phi_i\right)^2}{4\sum\limits_{j=1,2}
\Phi_j^\dag\Phi_j}\,,
\label{L57} 
\ee
where $\Delta{\cal L}$ has been chosen in the usual way. In calculating
the integral of ${\cal L}_{(5)}$ over $z$ one encounters three types of 
integrals,
$$
I_1=\int_{-\infty}^{\infty} dz\frac{\varphi_1^4(z)}{A \varphi_1^2(z)+B 
\varphi_2^2(z)}\,,\quad
I_2=\int_{-\infty}^{\infty} dz\frac{\varphi_2^4(z)}{A \varphi_1^2(z)+B 
\varphi_2^2(z)}\,, $$
\be
I_3=\int_{-\infty}^{\infty}dz\frac{\varphi_1^2(z)\varphi_2^2(z)}
{A \varphi_1^2(z)+B \varphi_2^2(z)}\,,
\label{I123}
\ee
where 
\be
A\equiv \phi_1(x)^\dag\phi_1(x)\,,\quad\quad B\equiv 
\phi_2(x)^\dag\phi_2(x)\,.
\label{AB}
\ee
Out of these three integrals, only one is independent. For example, one
can express $I_1$ and $I_2$ through $I_3$: 
\be
I_1=\frac{1-B I_3}{A}\,,\quad\quad  I_2=\frac{1-A I_3}{B}\,.
\label{rel}
\ee
Using these relations it is straightforward to check that the corresponding 
four-dimensional theory is gauge invariant.

\section{Discussion and conclusion}

We have proposed a simple mechanism of localization of gauge fields on a
brane which can work in space-times with an arbitrary number of extra 
dimensions, both flat and warped. The gauge fields are assumed to be  
composites made out of zero-mode matter fields localized on the brane. 
The localized matter fields may acquire masses through a mechanism different 
from the localization one; this would not destroy gauge invariance of the 
resulting vector field theory. 

We have considered several simple scalar and scalar-fermion models 
in which gauge bosons are dynamically generated, their kinetic terms 
being produced by quantum fluctuations of the localized matter fields. 
The mechanism is operative in both Abelian and non-Abelian cases. While pure 
fermionic models have difficulties ensuring gauge invariance, in models
with scalars exact gauge invariance can be naturally implemented. We 
demonstrated that the higher-dimensional gauge invariance translates into
the exact gauge invariance of the effective four-dimensional theory 
irrespective of the details of the localization mechanism of matter fields. 
Charge universality of gauge interactions is thus automatically preserved 
in the four-dimensional theory. One can expect that a similar mechanism
can also localize gravity on a brane.

\vspace{0.25cm} \noindent{\it Acknowledgements.} 
The author is grateful to V.A. Rubakov for very useful discussions. This 
work was supported by the ``Sonderforschungsbereich 375 f{\"u}r 
Astro-Teilchenphysik der Deutschen Forschungsgemeinschaft''. 


\end{document}